\documentclass[twocolumn,showpacs,amsmath,amssymb,superscriptaddress]{revtex4-2} %,groupedaddress
\usepackage{dcolumn,braket,color,amsmath,footnote,hyperref,bm}
\usepackage{graphicx,bm,url,longtable}
\usepackage{tabularx,multirow,braket}
\usepackage{fancybox}
\usepackage{tikz}
\usetikzlibrary{matrix}

\newcommand\+{\dagger}

\begin{document}

\title{Coupling of pairing and triaxial shape vibrations in collective
states of $\gamma$-soft nuclei}

\author{K.~Nomura}
\affiliation{Department of Physics, Faculty of Science, University of
Zagreb, HR-10000 Zagreb, Croatia}
\email{knomura@phy.hr}
\author{D.~Vretenar}
\affiliation{Department of Physics, Faculty of Science, University of
Zagreb, HR-10000 Zagreb, Croatia}
 \affiliation{ State Key Laboratory of Nuclear Physics and Technology, School of Physics, Peking University, Beijing 100871, China}
\author{Z.~P.~Li}
\affiliation{School of Physical Science and Technology, Southwest
University, Chongqing 400715, China}

\author{J.~Xiang}
\affiliation{School of Physics and Electronic, Qiannan Normal University
for Nationalities, Duyun 558000, China}
\affiliation{School of Physical Science and Technology, Southwest
University, Chongqing 400715, China}

\date{\today}

\begin{abstract}
In addition to shape oscillations, low-energy excitation spectra of deformed nuclei are also influenced by pairing vibrations. 
The simultaneous description of these collective modes and their coupling has been a long-standing problem in nuclear structure theory. 
Here we address the problem in terms of self-consistent mean-field calculations of collective deformation energy surfaces, and the framework of 
the interacting boson approximation. In addition to quadrupole shape vibrations and rotations, the explicit 
coupling to pairing vibrations is taken into account by a boson-number non-conserving Hamiltonian, specified by a choice of a 
universal density functional and pairing interaction. An illustrative calculation for $^{128}$Xe and $^{130}$Xe shows the importance of 
dynamical pairing degrees of freedom, especially for structures built on low-energy $0^+$ excited states, in $\gamma$-soft and triaxial nuclei.
\end{abstract}

\maketitle

\section{Introduction}
An accurate description of the structure of deformed nuclei that cannot be characterized by axially symmetric shapes presents 
a challenge for low-energy nuclear theory \cite{BM_II,RS}. Quadrupole shape deformations, in particular, can be described in terms of the polar 
variables $\beta$ and $\gamma$. The axial variable $\beta$ is
proportional to the intrinsic quadrupole moment, 
and the angular variable $0 < \gamma < \pi/3$ specifies the degree of triaxiality. Two limiting cases for non-axial nuclei correspond to: (i) a collective 
potential with a stable minimum at a particular value of $\gamma$
(the rigid-triaxial rotor model of Davydov and Filippov \cite{Davydov58})
and, (ii) a collective potential that is virtually independent of the
angular variable (the $\gamma$-unstable rotor model of Wilets and Jean
\cite{gsoft}). Numerous studies of the emergence of $\gamma$-softness
have shown that neither of the two limiting geometrical pictures is
realized in actual nuclei. Most non-axial medium-heavy and heavy nuclei
lie in between the limits of rigid-triaxiality and $\gamma$-unstable
rotors \cite{zamfir1991,CasBook,mccutchan2007,nomura2012tri,niksic2014}. 

An additional level of complexity is introduced by considering dynamical
pairing in addition to shape collective degrees of freedom
\cite{vaquero2013,giuliani2014,zhao2016,schunck2016,bernard2019}. 
The interplay between pairing and triaxial quadrupole deformations has
been a central subject in nuclear structure since the 1960s
\cite{bes1963,bes1966b,casten1972,ragnarsson1976}. 
The effect of coupling between shape and pairing vibrations is evident
in the excitation spectra, especially in the energies of
bands based on excited $0^+$ states, and the $E0$ transition
strengths \cite{BM_II,bes1966,bes1970,bes1972,brink-broglia,garrett2016}. The
dynamical pairing degree of freedom has been taken into 
account schematically in a number of studies that, however, did not
explicitly consider the coupling between shape and pairing vibrations
(see, for instances,
Refs.~\cite{prochniak1999,srebrny2006,prochniak2007}). In two recent
articles we have 
extended the quadrupole collective Hamiltonian \cite{xiang2020}, and the
interacting boson model (IBM) \cite{nomura2020pv}, to include pairing
vibrations and the coupling between shape and pairing degrees of
freedom. It has been shown that the coupling to pairing vibrations 
produces low-energy spectra in much better agreement with experimental
results. Both studies, however, have been restricted to axially
symmetric shapes. As noted in Ref.~\cite{xiang2020}, the effect of
pairing vibrations will particularly be important for $\gamma$-soft
nuclei characterized by shape coexistence \cite{heyde2011} and, therefore, it is
important to develop a model that allows for the coupling between
pairing and triaxial $(\beta, \gamma)$ shape degrees of freedom. In this
work we develop such a model based on nuclear density functional
theory and the IBM, and report the first microscopic calculation of
pairing and triaxial shape vibrations in collective states of
$\gamma$-soft nuclei. 
 
\section{Method}
To map the energy of a nucleus as function of intrinsic deformations, 
constrained self-consistent mean-field (SCMF)
calculations \cite{RS,bender2003,vretenar2005,niksic2011,robledo2019} are performed for a specific choice of the 
universal energy density functional and pairing force.
In this work we employ the self-consistent 
relativistic mean-field plus BCS (RMF+BCS) model \cite{xiang2012}, based on the 
density functional PC-PK1 \cite{PCPK1} and a separable pairing interaction \cite{tian2009}.
The constraints imposed in the present SCMF calculation 
are the expectation values of the quadrupole moments $\hat{Q}_{20}$ and $\hat{Q}_{22}$, 
and the monopole pairing operator $\hat{P}$. 
The expectation values of $\hat{Q}_{20} = 2z^2 - x^2 - y^2$ and
$\hat{Q}_{22} = x^2 - y^2$ determine the 
deformation parameters $\beta$ and $\gamma$, respectively. 
%dimensionless polar deformation parameters $\beta$ and $\gamma$. 
%that specify the elongation
%of a nucleus along its symmetry axis ($\beta$), and the degree of
%triaxiality ($\gamma$), respectively. 
The expectation value of the monopole pairing operator 
$\hat{P}={1}/{2}\sum_{k>0}(c_{k}c_{\bar{k}}+c^\+_{\bar k}c^\+_{k})$  
in a BCS state, where $k$ and ${\bar{k}}$ denote the single-nucleon and
the corresponding time-reversed states, respectively, 
defines the intrinsic pairing deformation parameter $\alpha$,
which can be related to the pairing gap $\Delta$. 
To reduce the computational complexity, no distinction is made 
between proton and neutron pairing degrees of
freedom even though, in principle, they should be treated separately.

\begin{figure*}[htb!]
\begin{center}
\begin{tabular}{cc}
 \includegraphics[width=0.5\linewidth]{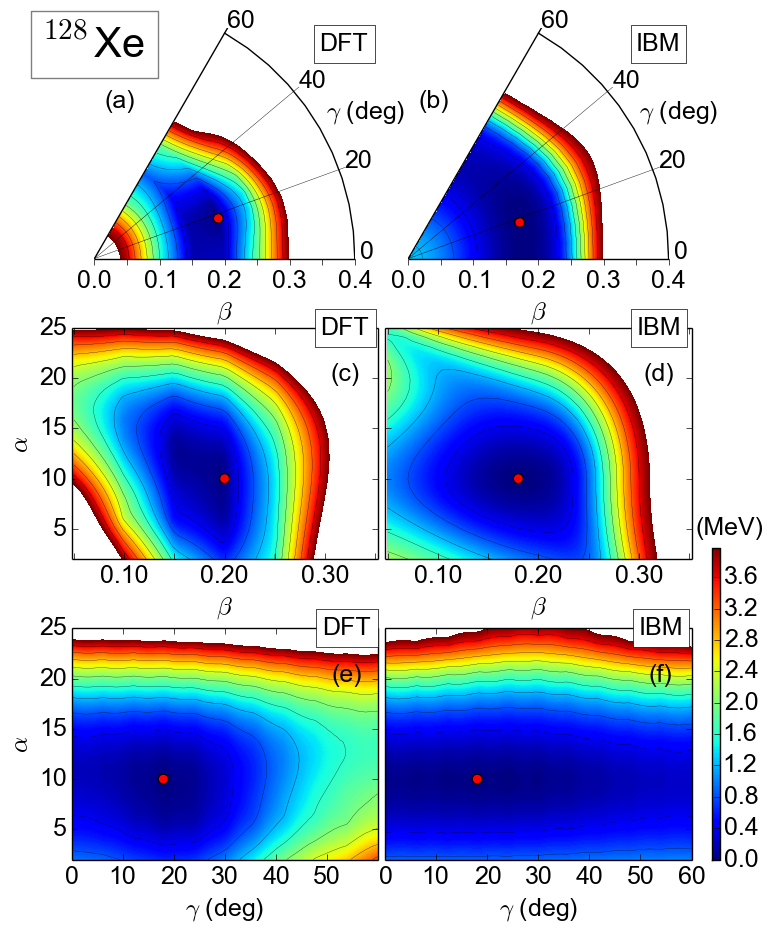} & 
\includegraphics[width=0.5\linewidth]{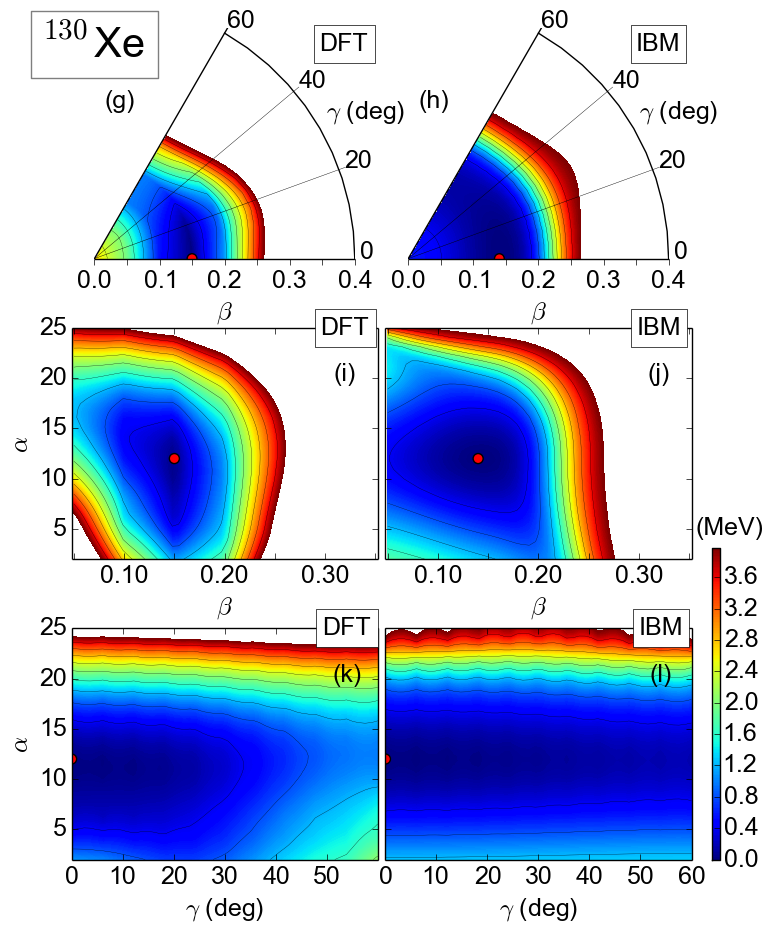}\\
\end{tabular}
\caption{Potential energy surfaces of $^{128,130}$Xe 
computed using the RMF+BCS model based on the functional 
PC-PK1 and a separable pairing interaction, and the interacting boson model (IBM) 
Hamiltonian determined by the microscopic SCMF energy maps (see text for the 
description). The 2D projections of the PESs are shown as functions of the axial quadrupole and
triaxial $(\beta,\gamma)$, axial quadrupole and pairing
$(\beta,\alpha)$, and triaxial quadrupole and pairing ($\gamma,\alpha$) deformations. 
The fixed values of $\alpha=10$ $(12)$ in the $(\beta,\gamma)$, $\gamma=18^\circ$ ($0^\circ$) in the 
$(\beta,\alpha)$, and $\beta=0.2$ (0.15) in the ($\gamma,\alpha$) plot, correspond to the global 
minimum of the $(\alpha,\beta,\gamma)$ PES of 
$^{128}$Xe ($^{130}$Xe).}
\label{fig:pes}
\end{center}
\end{figure*}

The three-dimensional potential energy surfaces (PES) of $^{128}$Xe and $^{130}$Xe, obtained 
in the $(\alpha,\beta,\gamma)$-constrained microscopic SCMF calculation, are projected 
onto two-dimensional planes in the first and third column of Fig.~\ref{fig:pes}, respectively. 
The PESs are plotted as functions of the axial quadrupole and
triaxial $(\beta,\gamma)$, axial quadrupole and pairing
$(\beta,\alpha)$, and triaxial quadrupole and pairing ($\gamma,\alpha$) deformations. 
The fixed values of $\alpha=10$ $(12)$ in the $(\beta,\gamma)$ plot, $\gamma=18^\circ$ ($0^\circ$) for the 
$(\beta,\alpha)$ surface and, finally, $\beta=0.2$ (0.15) in the ($\gamma,\alpha$) map, correspond to the global 
minimum in the entire $(\alpha,\beta,\gamma)$ parameter space of 
$^{128}$Xe ($^{130}$Xe). While both nuclei appear to be $\gamma$-soft (first row of Fig.~\ref{fig:pes}),  
the SCMF-$(\beta,\gamma)$ PES of $^{128}$Xe actually displays a shallow triaxial minimum at
$\gamma=18^\circ$. For $\gamma=18^\circ$ ($0^\circ$), the 
$(\alpha,\beta)$ surfaces of $^{128}$Xe ($^{130}$Xe) exhibit shallow minima at 
$\alpha=10$ $(12)$, respectively, and are rather soft with respect to 
the intrinsic pairing deformation parameter. As one can already infer from the first two 
maps, the $(\gamma,\alpha)$- energy surfaces at the minimum $\beta$ are soft with 
respect to both collective coordinates. Softness, of course, implies large fluctuations and, 
therefore, both the triaxial $\gamma$ and pairing $\alpha$ degrees of freedom will be important 
for spectroscopic properties of these two nuclei. 

%-----------------------------------------------------------
%
%	SCMF QP-PES -- Xe-128, 130
%
%-----------------------------------------------------------

To calculate excitation spectra and transition rates, one must 
extend the mean-field framework to include dynamical correlations that
arise from restoration of broken symmetries and fluctuations of
collective coordinates \cite{RS}. 
Physical quantities determined by collective dynamics are here computed by
mapping the SCMF results onto a system of interacting bosons \cite{nomura2008}. 
The boson model space consists of the monopole $s$ and
quadrupole $d$ bosons that are associated with correlated $J=0^+$
and $2^+$ pairs of valence nucleons, respectively. 
To take into account pairing vibrations, the number of bosons $n_0$ 
which, by construction equals half the number of valence nucleons \cite{OAI}, is not
conserved. Here we use a model with a  
Hilbert space expressed as a direct sum of three subspaces 
comprising  $n=n_0-1$, $n_0$, and $n_0+1$ bosons: 
\begin{align}
\label{eq:space}
 (sd)^{n_0-1}\oplus(sd)^{n_0}\oplus(sd)^{n_0+1}. 
\end{align}
The IBM Hamiltonian in general consists of boson-number conserving $\hat{H}_\mathrm{cons}$
and non-conserving $\hat{H}_\mathrm{non-cons}$ interactions:
\begin{align}
 \label{eq:ham}
\hat{H}_\mathrm{IBM} = \hat{H}_\mathrm{cons} + \hat{H}_\mathrm{non-cons}. 
\end{align}
%\begin{align}
% \label{eq:ham2}
%H = \epsilon_0 n + \epsilon_d n_d + \kappa Q\cdot Q
% + \rho L\cdot L + \eta{}T^\+\cdot\tilde T + \theta{}\Pi
%\end{align}
For a quantitative description of $\gamma$-soft nuclei, 
the Hamiltonian must include not only one- and 
two-body boson terms, but also three-body terms
\cite{vanisacker1981,heyde1984,casten1985,zamfir1991,nomura2012tri}. 
It has been shown that already a minimal choice for a three-body boson
interaction of the type $(d^{\+}d^{\+}d^{\+})^{(3)}\cdot (\tilde d\tilde d\tilde d)^{(3)}$, 
%\begin{align}
%(d^\+ d^\+ d^\+)^{(3)}\cdot (\tilde d\tilde d\tilde d)^{(3)}
%\end{align}
produces a triaxial minimum 
on the deformation energy surface and provides a correct description of 
the structure of $\gamma$-bands \cite{heyde1984,casten1985,zamfir1991,nomura2012tri}. 
The boson number non-conserving Hamiltonian $\hat{H}_\mathrm{non-cons}$ is
expressed in terms of a monopole pair-transfer operator $(s^\+ + s)$, that either adds or
removes an $s$ boson \cite{nomura2020pv}.

The bosonic PES in the $(\alpha,\beta,\gamma)$ space is computed 
by taking the expectation value of the IBM Hamiltonian in the boson condensate 
state $\ket{\Psi(\vec\alpha)}$
\cite{ginocchio1980,dieperink1980}: 
\begin{align}
 \ket{\Psi({\vec\alpha})}=\ket{\Psi_{n_0-1}({\vec\alpha})}\oplus\ket{\Psi_{n_0}({\vec\alpha})}\oplus\ket{\Psi_{n_0+1}({\vec\alpha})}, 
\end{align}
where, for a given subspace comprising $n$ bosons ($n=n_0-1,n_0,n_0+1$), 
$\ket{\Psi_{n}(\vec{\alpha})}$ is defined by 
\begin{align}
\label{eq:coherent}
\ket{\Psi_{n}({\vec\alpha})}=\Biggl[\alpha_s
 s^\+ + \sum_{m=-2}^{+2}\alpha_{m}d_{m}^\+\Biggr]^n\ket{0},
\end{align}
up to a normalization factor. The vector $\vec{\alpha}$ denotes the amplitudes $\alpha_s$ and
$\alpha_{m}$, and $\ket{0}$ is the boson vacuum. 
The bosonic energy surface is expressed as a $3\times 3$ matrix
${\bf E}(\vec{\alpha})$ \cite{frank2004}:
\begin{align}
 E_{n,n'}(\vec{\alpha})= 
&\braket{\Psi_n(\vec{\alpha})|\hat{H}_{\mathrm{cons}}|\Psi_n(\vec{\alpha})}
 \delta_{n,n'} \nonumber \\
&+
\braket{\Psi_{n'}(\vec{\alpha})|\hat{H}_{\mathrm{non-cons}}|\Psi_n(\vec{\alpha})}
 \delta_{n,n'\pm 1},
\end{align}
with the three indices $n_0$ and $n_0\pm 1$. 
%, with diagonal elements
%$E_{n,n}(\vec{\alpha})=\braket{\Psi_n(\vec{\alpha})|H_{\mathrm{consrv}}|\Psi_n(\vec{\alpha})}$
%$(n=n_0-1,n_0,n_0+1)$ and non-diagonal ones 
%$E_{n,n'}(\vec{\alpha})=\braket{\Psi_{n'}(\vec{\alpha})|H_{\mathrm{non-consrv}}|\Psi_n(\vec{\alpha})}$
%($n'= n\pm 1$).

The amplitudes $\alpha_s$ and $\alpha_{m}$ in \ref{eq:coherent} can be related 
to the pairing and triaxial deformation parameters of the SCMF
calculations, respectively \cite{nomura2020pv}. 
The boson Hamiltonian (\ref{eq:ham}) is determined using the method of Ref.~\cite{nomura2020pv}.  
The parameters of the boson number conserving Hamiltonian
$\hat{H}_{\mathrm{cons}}$ are 
specified by mapping the $(\beta,\gamma)$-SCMF PES at 
$\alpha=\alpha_\mathrm{min}$ onto the diagonal matrix element 
$E_{n_0,n_0}(\alpha=\alpha_\mathrm{min},\beta,\gamma)$. We 
note that only the strength of the rotational term $\hat{L}\cdot\hat{L}$
in $\hat{H}_\mathrm{cons}$ is determined 
separately \cite{nomura2011rot}, by adjusting the 
moment of inertia of the yrast band to the empirical value. For 
$^{128,130}$Xe this value is $\approx 40 \%$ larger than 
the corresponding Inglis-Belyaev value \cite{inglis1956,belyaev1961}, 
computed using the SCMF single-nucleon quasiparticle states at the 
equilibrium minimum.  
The strength parameter of the number non-conserving Hamiltonian
$\hat{H}_{\mathrm{non-cons}}$ is chosen in such a way that the
$(\alpha,\beta)$-SCMF PES at 
$\gamma=\gamma_\mathrm{min}$ is reproduced 
by the lowest eigenvalue of the matrix ${\bf E}(\vec{\alpha})$. The details of the formalism for the 2D space 
$(\alpha,\beta)$ in the case of axial symmetry can be found in Ref.~\cite{nomura2020pv}, and the expressions 
used in the extension to the triaxial case will be included in a forthcoming publication.

The three projections of the IBM PESs on the $(\beta,\gamma)$, $(\alpha,\beta)$, and
$(\gamma,\alpha)$ planes are shown in the second and fourth column of Fig.~\ref{fig:pes} 
for $^{128}$Xe and $^{130}$Xe, respectively. They are displayed next to the corresponding 
microscopic energy surfaces so that one can assess 
the mapping from the SCMF space of nucleon degrees of freedom to the boson space of the IBM. 

We note that in a more traditional microscopic approach to large-amplitude collective motion, 
such as the collective Hamiltonian model \cite{prochniak2009,niksic2011}, the dynamics is governed by 
the collective potential, the mass parameters, and moments of inertia, all defined as functions of the 
intrinsic deformation parameters. The single-nucleon wave functions, energies and occupation factors, 
generated from constrained SCMF calculations, provide the microscopic input for the parameters of the 
collective Hamiltonian. In the present approach, the collective dynamics is determined by the choice of 
the boson space ($s$ and $d$ bosons) and the IBM Hamiltonian that includes not only one-body, but 
also two-body and three-body boson interaction terms. Even though the parameters of this Hamiltonian 
do not explicitly depend on the intrinsic deformation parameters, the mapping of the entire SCMF 
energy surface on the expectation value of the IBM Hamiltonian in the boson condensate state, 
introduces an effective deformation dependence of the boson Hamiltonian. Of course, 
at very large deformations intruder orbitals become important, and the mapping to the limited boson 
space that corresponds to half the number of valence nucleons is too restrictive. However, in the 
vicinity of the equilibrium minimum the mapping is quite accurate (c.f. Fig.~\ref{fig:pes}), and 
generally produces a boson Hamiltonian that can describe low-energy excitation spectra at a 
quantitative level. 

%-----------------------------------------------------------
%
%	Systematics - 0+
%
%-----------------------------------------------------------
\begin{figure}[htb!]
\begin{center}
\includegraphics[width=.7\linewidth]{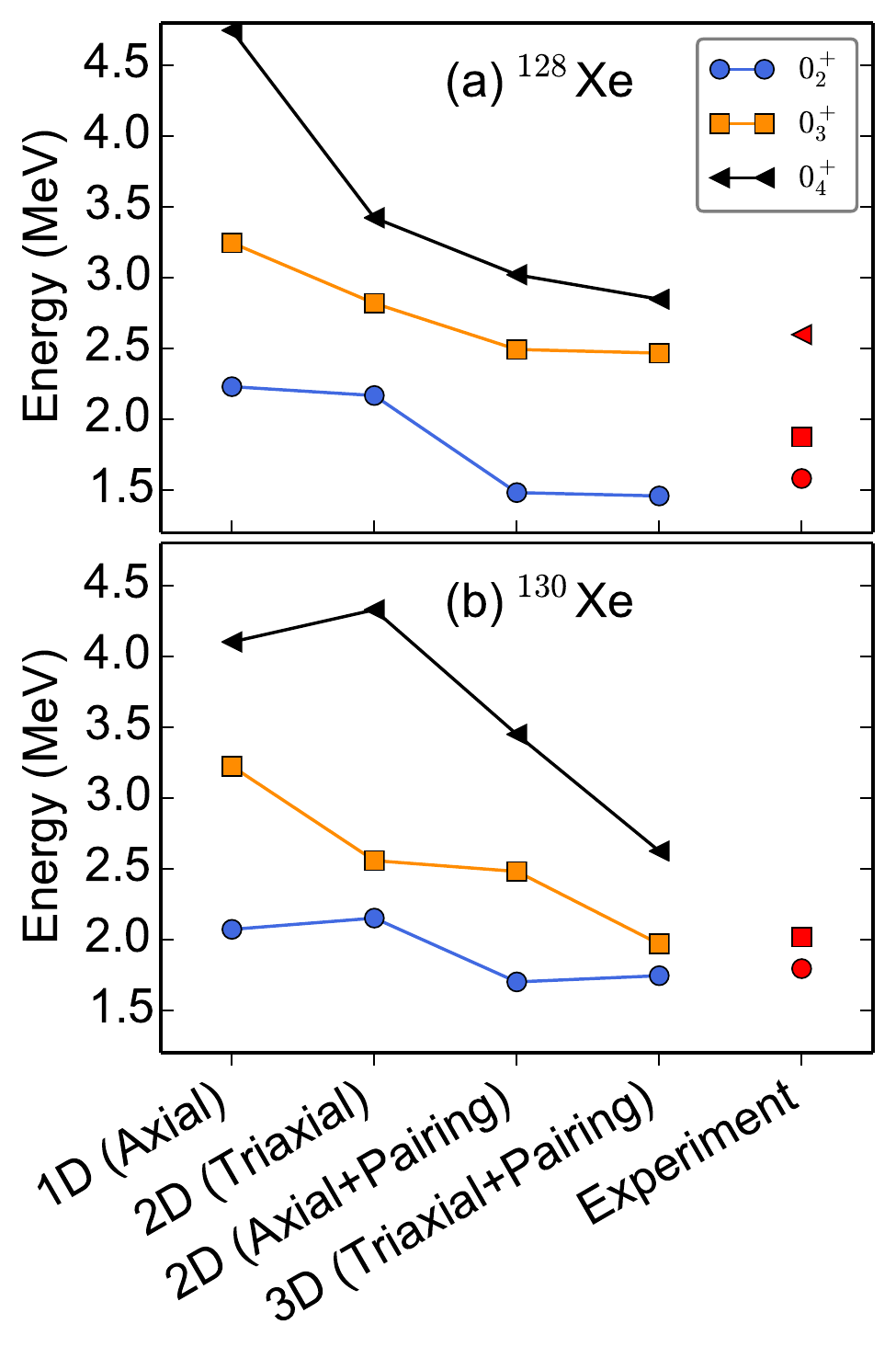}
\caption{Excitation energies of the second, third, and
 fourth $0^+$ states in $^{128}$Xe and $^{130}$Xe. 
Results obtained with the IBM including 1D axial quadrupole ($\beta$), 2D triaxial
 quadrupole ($\beta,\gamma$) and pairing plus axial quadrupole
 ($\alpha,\beta$), and 3D pairing plus triaxial quadrupole
 ($\alpha,\beta,\gamma$) degrees of freedom are compared with experimental 
 values (filled red symbols on the
 right-hand side of each panel).}
\label{fig:sys_0+}
\end{center}
\end{figure}

%-----------------------------------------------------------
%
%	Systematics - gamma
%
%-----------------------------------------------------------
\begin{figure}[htb!]
\begin{center}
\includegraphics[width=.9\linewidth]{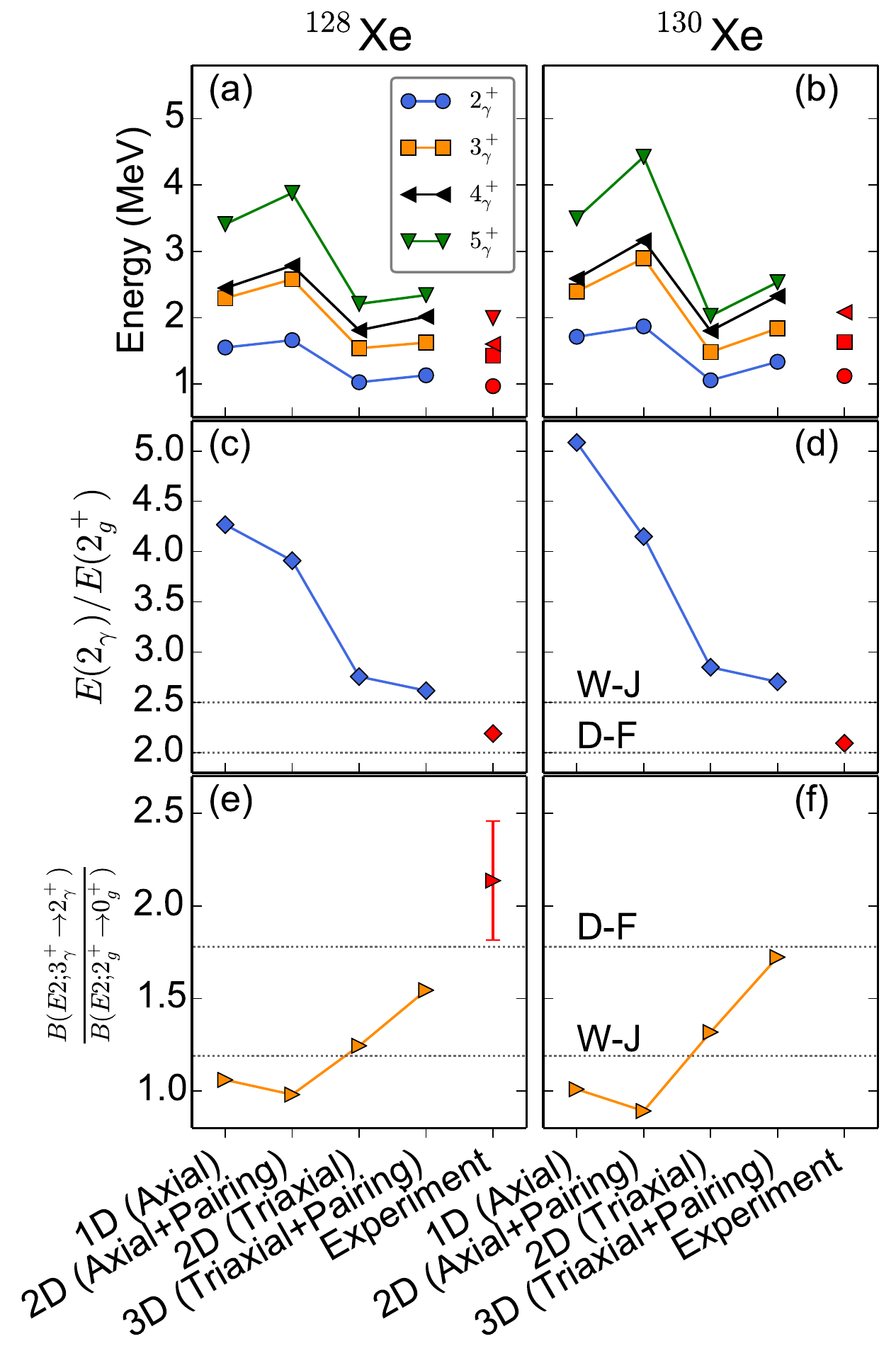}
\caption{Excitation spectra of the
 states belonging to the $\gamma$-band (a,b), the energy ratio
 $E_{2\gamma} = E(2^+_\gamma)/E(2^+_g)$ (c,d), and the ratio 
 $R_{3\gamma}=B(E2;3^+_\gamma\to 2^+_\gamma)/B(E2; 2^+_g\to 0^+_g)$
 (e,f), obtained by four different IBM calculations 
of $^{128}$Xe and $^{130}$Xe:
 axially deformed, axially deformed + dynamical pairing, triaxially deformed, and triaxially deformed + dynamical pairing. 
 In panels (c,d,e,f), the values predicted 
by the triaxial rotor model of Davydov and 
 Filippov (D-F) at $\gamma=30^{\circ}$ \cite{Davydov58} ($E_{2\gamma}=2.00$
 and $R_{3\gamma}=1.78$), and by the $\gamma$-unstable-rotor
 model of Wilets and Jean (W-J) or O(6) symmetry \cite{gsoft,IBM} ($E_{2\gamma}=2.50$
 and $R_{3\gamma}=1.19$) are also indicated by dotted horizontal lines. Available experimental 
 values from Refs.~\cite{data,coquard2009,peters2016} are shown 
by filled red symbols on the right-hand side of each panel. }
\label{fig:sys_xe130}
\end{center}
\end{figure}

%Note the error for the experimental $R_{3\gamma}$ ratio is not plotted in the figure, since that of
%the $B(E2;3^+_\gamma\to 2^+_\gamma)$ transition rate is too large (see Fig.~\ref{fig:level}).

%-----------------------------------------------------------
%
%	Spectra - Xe-128,130
%
%-----------------------------------------------------------
\begin{figure}[htb!]
\begin{center}
\includegraphics[width=\linewidth]{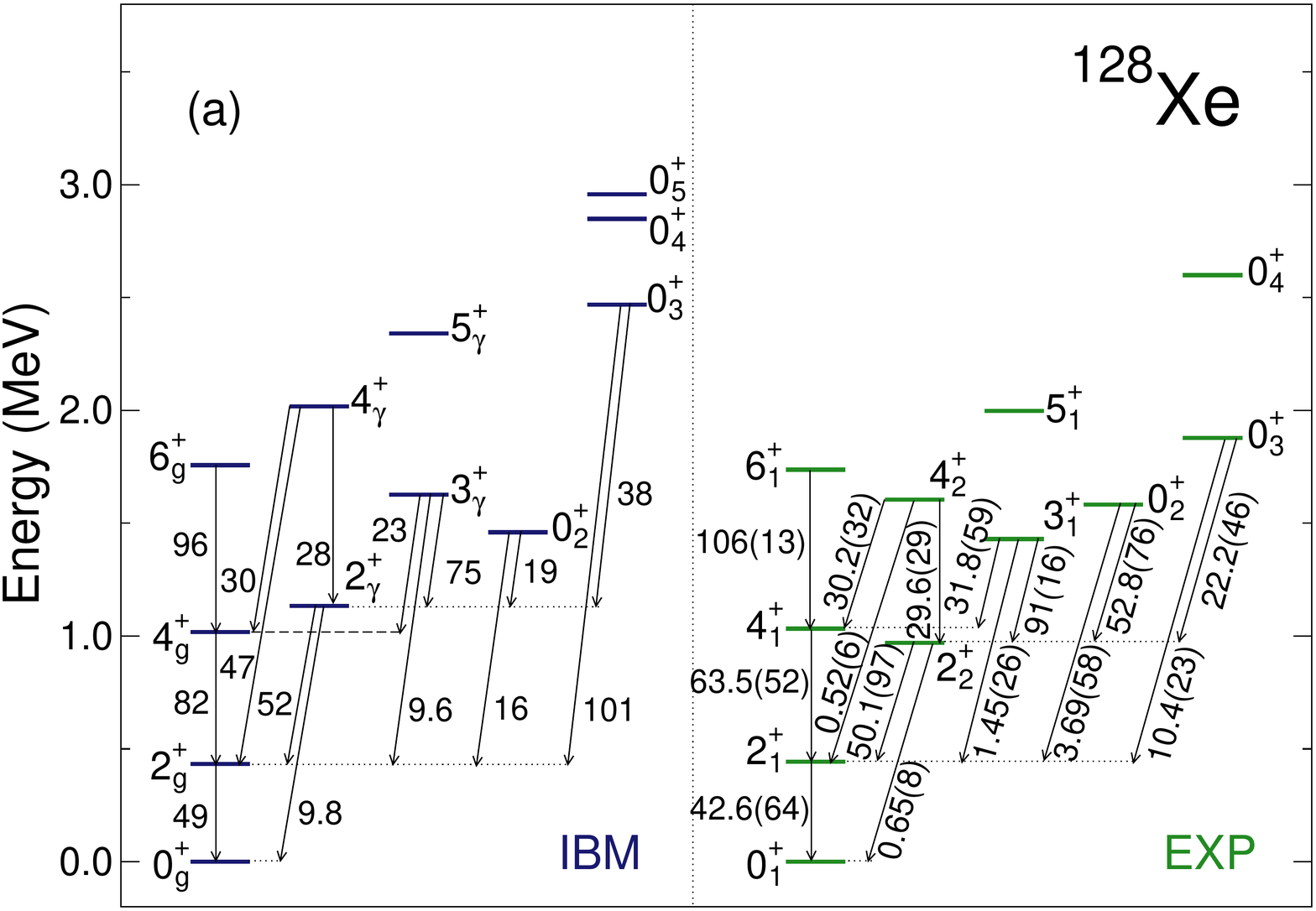} \\
\includegraphics[width=\linewidth]{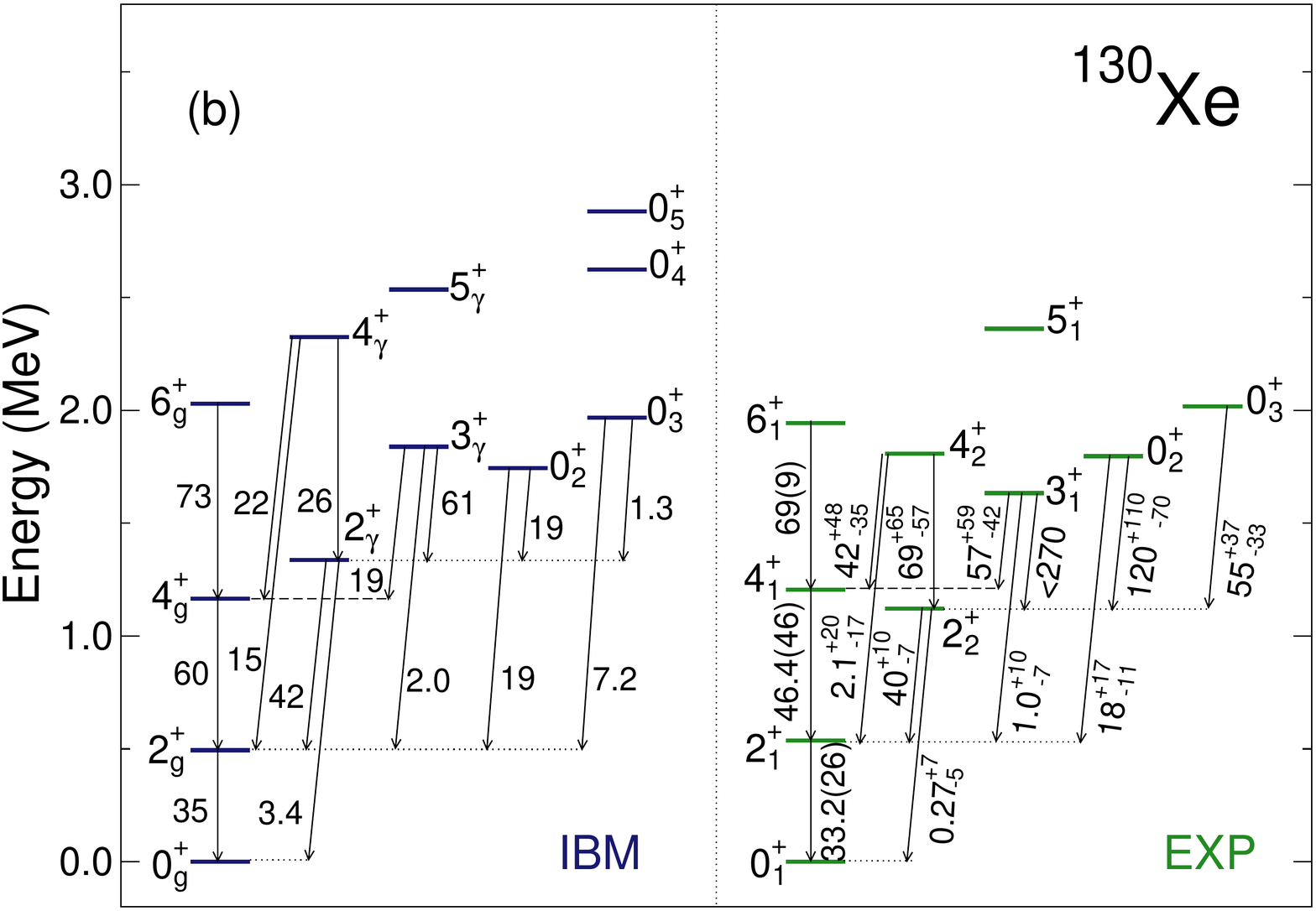} 
\caption{Low-energy excitation spectra of $^{128,130}$Xe 
 obtained with the IBM that includes dynamical pairing and triaxial 
 deformation degrees of freedom. The levels are grouped into bands according 
 to the dominant transitions, and the $B(E2)$ values are in Weisskopf
 units. The results of model calculation are compared to the corresponding experimental energy spectra 
 \cite{data,coquard2009,peters2016}.} 
\label{fig:level}
\end{center}
\end{figure}

\section{Effect of dynamical pairing and triaxial deformation on excitation spectra}
Having determined the parameters of the IBM Hamiltonian, we next 
consider spectroscopic properties and discuss the
importance of simultaneously incorporating dynamical pairing and
triaxial degrees of freedom in the model space. 
As already shown in Ref.~\cite{nomura2020pv} for axially symmetric calculations of 
$^{122}$Xe and rare-earth $N=92$ isotones, the coupling between shape and pairing collective degrees
of freedom has hardly any effect on states of the yrast band, either on excitation energies or transition rates. 
In contrast, the inclusion of dynamical pairing significantly lowers the energies of bands based on excited $0^+$ states.
Figure \ref{fig:sys_0+} displays the excitation energies of the second, third, and
 fourth $0^+$ states in $^{128}$Xe and $^{130}$Xe. We plot 
the energies calculated with the IBM including 1D axial quadrupole ($\beta$), 2D triaxial
 quadrupole ($\beta,\gamma$) and pairing plus axial quadrupole
 ($\alpha,\beta$) and, finally, 3D pairing plus triaxial quadrupole
 ($\alpha,\beta,\gamma$) degrees of freedom. The lines are a guide to the eye, and the 
 experimental values are denoted by filled red symbols on the
 right-hand side of each panel. Because of configuration mixing it is not possible  
 to uniquely separate the effects of triaxial deformations and pairing vibrations on each 
$0^+$ state. However, the inclusion of these degrees of freedom generally 
lowers the $0^+$ states, bringing the excitation energies in
a quantitatively better agreement with 
experiment. In the particular examples considered here, it appears that the energies of 
$0^+_2$ are not sensitive to the inclusion of triaxial deformations, whereas both 
$\gamma$-deformation and dynamical pairing have an effect on the 
excitation energies of $0^+_3$ and $0^+_4$. 
One should keep in mind that these model calculations are performed in the 
collective monopole and quadrupole boson space. In actual nuclei, however, 
two- or four-quasiparticle states play a role at higher excitation energies, 
e.g. above $\approx 3$ MeV, but these degrees of freedom are not included in 
our model space. Even though a mixing with these states would, of course, affect 
the calculated excitation energies, the qualitative effect of dynamical pairing would 
still be the lowering of excited $0^+$ states. 

A similar analysis is performed for the excitation energies of the members of the 
$\gamma$-bands of $^{128}$Xe and $^{130}$Xe, and illustrated in the top panels 
of Fig.~\ref{fig:sys_xe130}. As one would expect, in this case the effect of the 
inclusion of triaxial deformations is important to reproduce the experimental 
excitation energies of the members of the $\gamma$ band. One notes,
however, that the coupling with pairing vibrations  
increases the excitation energy of the $\gamma$-band. 
This is somewhat at variance with the experimental data, the deviation attributed
to the level repulsion due to the configuration mixing between the
subspaces (\ref{eq:space}) of the IBM.

We have further analyzed two quantities that characterize the level of 
$\gamma$-softness. The limiting cases are described by 
two geometrical models: the rigid-triaxial-rotor model
of Davydov and Filippov (D-F), and the $\gamma$-unstable rotor model of Wilets and
Jean (W-J). The latter is equivalent to the O(6) dynamical symmetry of
the IBM \cite{arima1979o6,IBM}. To distinguish between rigidity
and softness in $\gamma$, we consider the ratios
$E_{2\gamma}=E(2^+_\gamma)/E(2^+_g)$ and 
 $R_{3\gamma}=B(E2;3^+_\gamma\to 2^+_\gamma)/B(E2; 2^+_g\to 0^+_g)$. 
 They are plotted in the middle and lower rows of
 Fig.~\ref{fig:sys_xe130}, respectively.
Note that for the $B(E2)$ values, because three-body
boson terms are included in the Hamiltonian, calculations with triaxial degree of
freedom should in principle contain higher-order
terms in the $E2$ transition operator  \cite{heyde1984}. Both calculated 
quantities exhibit a pronounced dependence on the triaxial degree of 
freedom. The results of the full calculation for $E_{2\gamma}$ are closer to 
the W-J limit, while the $R_{3\gamma}$ values trend towards the D-F limit.

Finally, we demonstrate that the model is also capable of describing detailed
structure properties of $\gamma$-soft nuclei. In Fig.~\ref{fig:level} 
the low-energy excitation spectra of $^{128}$Xe and $^{130}$Xe,
obtained with the IBM that includes the dynamical pairing and triaxial 
deformation degrees of freedom, are compared to the corresponding
experimental energy spectra \cite{data,coquard2009,peters2016}. 
On closer inspection it is seen that the present IBM calculation
reproduces the available low-energy data. 
%based on the microscopic framework of nuclear energy density
%functionals and using a 3D space of collective coordinates that includes the
%coupling between triaxial shape and 
%pairing vibrations, 
Characteristic features of $\gamma$-soft nuclei emerge in
the calculated excitation spectra: 
the low energy of the band-head of 
$\gamma$ band ($2^+_\gamma$), the 
level spacing between the states of the $\gamma$ band, and the
excitation energy of the $0^+_2$ state relative to the $\gamma$ band.

%$Q_s=\sqrt{16\pi/5}\biggl(\begin{array}{ccc}
%2 & 2 & 2 \\
%-2 & 0 & 2
%     \end{array}\biggr)\braket{2^+_1 \| T^{(E2)}\| 2^+_1}=-0.0074$
%     $e\cdot$b. 
For $^{128}$Xe the present calculation predicts the $\gamma$-band at somewhat higher 
excitation energy compared to its experimental counterpart. The $0^+_2$
state is slightly lower than the corresponding experimental level,
while the calculated $0^+_3$ is at considerably higher excitation energy. The $0^+_2$ wave function is 
dominated by components that are almost equally distributed between the $[n_0-1]$ and $[n_0+1]$ subspaces (more than 90\%), 
that is, the structure corresponds to a pairing vibrational state. In contrast, the boson distribution of $0^+_3$ is very similar to that of the 
ground-state band, with predominant components in the $[n_0]$
subspace. This is reflected in the strong $E2$ transition 
to the state $2^+_1$ ($2^+_g$), with a $B(E2)$ value an order of magnitude larger than in 
experiment. The observed $E2$ transition strengths
\cite{coquard2009,data} are, in general, reproduced by the model 
calculation, except for the weak transition $4^+_2\to 2^+_1$. The
calculated $B(E2)$ value is, in fact, 
larger than that of the corresponding $4^+_\gamma \to 4^+_{g}$ transition,
which reproduces the experimental value.  

The calculated spectrum of $^{130}$Xe reproduces the data equally well
and, in fact, the excitation energies of  
the states $0^+_2$ and $0^+_3$ are in better agreement with experiment
compared to the previous case.  
Here both $0^+_2$ and $0^+_3$ exhibit a structure that can be
interpreted as pairing vibrations, while the  
largest component of the wave function of $0^+_4$ is that of the $[n_0]$
subspace. Just as in the case of $^{128}$Xe,  
the $\gamma$-band is calculated at somewhat higher excitation energy
with respect to experiment. The predicted transition rates are 
consistent with the available data even though, except for yrast band,
the latter are dominated by large error bars.

\section{Conclusion}
Based on self-consistent mean-field calculations of deformation energy surfaces, and the framework of the 
interacting boson approximation, a new method has been developed that allows for the coupling between pairing 
and triaxial $(\beta, \gamma)$ shape degrees of freedom. In addition to quadrupole shape vibrations and rotations, the explicit 
coupling to pairing vibrations is taken into account by a boson-number non-conserving IBM Hamiltonian. The parameters of the 
Hamiltonian are specified by SCMF calculations for a specific choice of a universal energy density functional and pairing interaction, with 
constraints on quadrupole shape and pairing intrinsic deformations.  
The illustrative calculation of low-energy excitation spectra of
$^{128}$Xe and $^{130}$Xe indicates the importance of the 
dynamical pairing degree of freedom, especially for low-energy $0^+$
excited states and bands based on them. The findings of the 
present study will pave the way for more detailed explorations of pairing 
vibrations in various regions of $\gamma$-soft and triaxial nuclei. 

\begin{acknowledgments}
 This work has been supported by the Tenure Track Pilot Programme of 
the Croatian Science Foundation and the 
\'Ecole Polytechnique F\'ed\'erale de Lausanne, and 
the Project TTP-2018-07-3554 Exotic Nuclear Structure and Dynamics, 
with funds of the Croatian-Swiss Research Programme. 
It has also been supported in part by the QuantiXLie Centre of
Excellence, a project co-financed by the Croatian Government and
European Union through the European Regional Development Fund - the
Competitiveness and Cohesion Operational Programme (KK.01.1.1.01). 
The author Z.P.L. acknowledges support by the NSFC under Grant
No. 11875225. The author J.X. acknowledges support by the NSFC under
Grants No. 12005109 and No. 11765015.
\end{acknowledgments}

\bibliography{refs}

\end{document}